\documentstyle[12pt,epsfig]{article}
\begin{document}

\vspace*{2.cm}
\begin{center}
{\bf Neutrino Spectrum Distortion  Due to Oscillations
 and its BBN Effect}   

\vspace*{0.2cm}
Daniela Kirilova$^{\dagger}$\\[0.3cm]
{\it $^\dagger$Institute of Astronomy, Sofia\\
and  Physique Theorique, ULB, Bruxelles}
\end{center}
\vspace*{0.2cm}
\begin{abstract}

We  study the  distortion of electron neutrino 
energy spectrum  due to oscillations with the sterile neutrino 
 $\nu_e\leftrightarrow \nu_s$,
 for 
different initial  populations of the  sterile state  $\delta N_s$ at 
the onset of oscillations. The influence of this spectrum distortion 
on Big Bang Nucleosynthesis is analyzed. 
 Only the case of an initially empty  sterile state was  studied in  
previous publications. 

The primordial abundance of He-$4$ is calculated  
 for  all possible $\delta N_s$:
 $0 \le \delta N_s \le 1$ in the model of oscillations, effective 
after electron neutrino decoupling, for which the spectrum distortion 
effects on the neutron--proton transitions are the strongest. 
 
    It is found that   the spectrum 
distortion effect may be  the dominant one  not only in the case of  
small $\delta N_s$,  but also in the case of  big initial population of 
 $\nu_s$. For example, in the resonant case it may   
play a considerable role even for very big  $\delta N_s\sim 0.8$. 

\ \\

keywords: cosmology, primordial nucleosynthesis, neutrino oscillations

\end{abstract}

\newpage 
\vspace*{0.5cm}
\section*{Introduction}

Sterile neutrinos $\nu_s$  
 may be present at the onset of 
the nucleosynthesis epoch.   There may be different 
reasons for their  production  --- they are 
naturally produced in GUT models~\cite{brahmachari},  in models with 
large extra 
dimensions~\cite{extradim} and Manyfold Universe models~\cite{Manyfold}, 
in mirror matter models~\cite{mirror}. 
They may be produced also in 
 $\nu_{\mu,\tau}\leftrightarrow \nu_s$ oscillations in the 
preceeding epoch~\footnote{For example, atmospheric and LSND neutrino data 
require oscillations with
maximal mixings and  mass
differences $\delta m^2_{atm}\sim 10^{-3}$ eV$^2$ and
$\delta m^2_{LSND}\sim 10^{-1}$ eV$^2$,  which are effective
before the  BBN epoch. So, in  many schemes owing  to  
$\nu_{\mu}\leftrightarrow \nu_s$
or  $\nu_{\tau}\leftrightarrow \nu_s$
oscillations,  $\nu_s$ state may be  partially thermalized
before the
nucleosynthesis epoch.} 
according to the most popular  4-neutrino mixing  
schemes~\cite{mixing}.

 The degree of population of $\nu_s$,  $\delta N_s$, may be different  
depending on the 
concrete model of $\nu_s$ production. Hence, we will consider further on 
$\delta N_s$ as a  parameter.  

In recent years, strong constraints on the sterile neutrino impact in 
oscillations, explaining atmospheric and solar neutrino anomalies, were
obtained from the analysis of experimental oscillations 
data~\cite{constraints,strumia,conca}.
And although  a pure sterile channel solution is
excluded for any of the neutrino anomalies,
the analysis  of neutrino oscillation experimental data allows a certain,
small fraction of sterile neutrino to
participate into these oscillations. In the analysis ~\cite{barger} and  
~\cite{Holanda} it was suggested that a small fraction of
$\nu_s$ is not only allowed but even desirable for
 solar neutrino data. Recent measurements  of the cross section of
$^7Be(p,\gamma)^8B$ reaction lead to a predicted total $^8B$ neutrino flux
by $13\%$ larger than measured by SNO, which also may be  providing evidence for
sterile neutrino~\cite{gai}.

There also exist stringent cosmological constraints on $\nu_s$ produced in 
oscillations, 
based on oscillations influence  on BBN nucleosynthesis of $^4\!He$. 
For the case  $\nu_{\mu,\tau}\leftrightarrow \nu_s$ see for 
example refs.~\cite{dol,Villante}  
and for the   $\nu_e\leftrightarrow \nu_s$  see refs.~\cite{NPS,Villante}.
The  constraints on electron--sterile
neutrino oscillations~\cite{res,PRD}, excluded the active--sterile LOW 
solution
to the solar neutrino puzzle, in addition to the already 
excluded in  pioneer works (see for example ref.\cite{bd}) sterile LMA 
solution and  sterile solution to the atmospheric 
neutrino anomaly.
(For more detail about constraints on neutrino oscillations from cosmology, 
see  refs.~\cite{dubnastro, dol}.)

Most of  the cosmological constraints on  active-sterile mixing 
were obtained  in simple two-neutrino mixing schemes
~\footnote{For discussion and calculation of  cosmological
constraints in a specific 4-neutrino mixing schemes see 
ref.~\cite{bilenky,DiBari,Villante}}, 
 for the case when  the sterile neutrino 
state was  initially empty  at the epoch before  oscillations    
became effective  in the Universe evolution, $\delta N_s=N_{\nu}-3=0$. 
$N_{\nu}$ is the number of 
neutrino species in equilibrium. 
Since  the presence of a non-empty sterile
state before  oscillations was not considered in previous
analysis of oscillations effects on the neutrino spectrum  
distortion and  on BBN, in this work we  address this question.
We  omit the  assumption   $\delta N_s=0$ and explore the  
general case   $\delta N_s \ne 0$. 

 In the  $\nu_{\mu,\tau}\leftrightarrow \nu_s$ oscillation  case, 
 $\delta N_s \ne 0$ present initially  
  just leads to an earlier increase of the total energy density of the 
Universe, and it is straightforward to re-scale the existing constraints. 
In the  $\nu_e\leftrightarrow \nu_s$ oscillations case, however, the 
presence of 
$\nu_s$ at the onset of oscillations influences also  
  the kinetic 
effects of  $\nu_e\leftrightarrow \nu_s$ on BBN. Therefore, we chose 
 $\nu_e\leftrightarrow \nu_s$  
oscillations case for  exploring   electron neutrino distortion, caused by  
oscillations,  
and its influence on nucleons freezing and on  primordial  $^4\!$He,  
$Y_p$ for different  $\delta N_s$ values in the range   
 $0 \le \delta N_s\le 1$.

\section*{The spectrum distortion of the electron neutrinos} 

We analyze the case of oscillations effective after neutrino
decoupling, therefore further on we  denote by  $\delta N_s$
the degree of population of the sterile neutrino state at
active neutrino decoupling ($T\sim 2$ MeV).
Here we  explore     $\nu_e$
spectrum distortion , 
considering the degree of population of the 
sterile neutrino state  $\delta N_s$ 
as a free parameter and, varying its value in the range $[0,1]$ with
a step $0.1$. 
In the next section   we 
calculate  the kinetic effect of oscillations on 
primordial 
abundance of  $^4\!$He for different  $\delta N_s$. 

The mixing of electron neutrino with  $\nu_s$ has the following two types of  
effects on BBN:\\
(a) it leads to  {\it an increase of the energy density} of the
Universe~\cite{d81}, and \\
(b) it {\it changes the nucleons  kinetics} essential  for
$n/p$-freezing,   through the depletion of electron 
neutrino number
density,  distortion of  the equilibrium spectrum of $\nu_e$, 
 and   production of
asymmetry
between neutrinos and anti-neutrinos, all due to oscillations~\cite{dubnastro}.
We will parametrize these kinetic effects and denote them  $\delta N_{kin}$ 
further on.

(a) The first effect is usually described by  an increase
of the effective
number of
the energy
density degrees
of freedom $g_{eff}=(30/\pi^2)(\rho/T^4)$. At the
BBN epoch  $g_{eff}=10.75+7/4\delta N_s(T_{\nu}/T)^4$.
 Hence, this effect leads to a faster expansion rate $H\sim g_{eff}^{1/2}$ 
and
higher freezing
temperature for nucleons $T_f\sim g_{eff}^{1/6}$, when nucleons were more 
abundant:
$$
n/p\sim \exp\left[-\Delta m/T_f\right]
$$
 
This  
reflects  into
an overproduction of  $^4\!$He, since it strongly depends on the $n/p$-freezing 
ratio: $Y_p\sim 2 \exp(-\Delta m/T_f)/[1+ \exp(-\Delta m/T_f)]$,
where $\Delta m=m_n-m_p\sim1.3$ MeV is the neutron--proton mass 
difference.
This  effect of $g_{eff}$ increase on  BBN is well known~\cite{g}. The
approximate fit to the exact
calculations is:
       $\delta Y_p \sim 0.013\delta N_s$. 
 The maximum helium overproduction corresponding to $\delta N_s=1$ 
is $\sim 5\%$.

(b) The influence of the kinetic  effects of  oscillations
 on BBN is quite obvious, having in mind that:
(i) oscillations take
place between
equilibrium electron neutrino and  less populated  sterile neutrino 
ensemble,
 that (ii) the oscillations probability is inversely  proportional to the
energy
of  neutrinos 
 $P \sim \delta m^2/E$, so that neutrinos with different momenta start oscillating
at different cosmic times, 
and that (iii) the proton density is bigger than the neutron one. 
 Due to
that,
the neutrino energy spectrum $n_{\nu}(E)$ may strongly deviate
from its equilibrium form~\cite{kir, dubnastro}. 

 In case oscillations proceed   after
the decoupling of active neutrinos,
a strong spectrum distortion for both  the electron neutrino and
the anti-neutrino is possible.
This spectrum distortion
  affects the kinetics of nucleons freezing - it leads to an earlier
$n/p$-freezing and an overproduction  of
 $^4\!$He  yield.

The effect  can be easily understood  having in mind that
the  distortion leads both to a {\it depletion of the active neutrino
number densities}  in favor of the sterile ones $N_{\nu}$:
$$
N_{\nu}\sim \int {\rm d}E E^2 n_{\nu}(E)
$$

\noindent and to a {\it decrease of the mean neutrino energy}.~\footnote{
The decrease of the electron neutrino energy due to
oscillations into low temperature sterile neutrinos,
has also an additional effect: Due to the threshold of
the reaction converting protons into neutrons,
when neutrinos have lower energy than the threshold one,
 protons are preferably produced,
which may lead to an under-production of  $^4\!$He~\cite{dk}.
However, this turns to be a minor effect in the discussed oscillations 
model.}
This lowers the weak rates, governing nucleons transitions during
neutrons freezing, with respect to their values in the standard BBN model,
$\Gamma_{weak}\sim N_{\nu_e} \bar E_{\nu}^2$, and hence reflects into
 earlier freezing when neutrons were more abundant.
So, helium is   
over-produced.

The generation of neutrino-antineutrino asymmetry in the resonant oscillations 
has a subdominant effect on  $^4\!$He. It slightly suppresses oscillations at small 
mass differences, leading to a decrease of helium overproduction.

 $\delta N_{kin}$  depend strongly on the initial population  of the
sterile neutrino at BBN.
Larger  $\delta N_s$  decreases the kinetic  
effects,
because the element of initial non-equilibrium between the active and the
sterile states is less expressed.
Hence, for any specific value of  $\delta N_s$
it is necessary to provide a separate analysis.

In the case  $\delta N_s=1$~~~ $\nu_s$ are in
equilibrium (the sterile state is
as abundant as the electron one),   and hence 
the $n$--$p$ 
kinetics does not feel the oscillations,  $\delta N_{kin}=0$.  
 The final effect is  only due
to the energy increase, i.e.  $\delta N_{tot}=1$.

In  the case $\delta N_s=0$ the kinetic effect of oscillations  was
studied 
numerically 
for both  the resonant~\cite{res}  
 and non-resonant~\cite{PRD} oscillation cases.
In this case   $\delta N_{kin}$  for given fixed
mixing parameters
reach their highest value,  $\delta N_{kin}^{max}$, as far as the non-equilibrium
element  --- the difference
between the sterile and active neutrino number densities at the beginning
of
oscillations --- is the greatest.
 The overproduction of $^4\!$He may be enormous:  up to $13.2\%$ in the
non-resonant oscillation case and up to $31.8\%$ in the
resonant one~\cite{bern}.
This corresponds effectively to a little more than $6$ additional neutrino 
states
 $\delta N_{kin}^{max}\sim 6$.
 For  the case   $\delta N_s=0$ the kinetic effects were carefully studied
  and  stringent cosmological constraints on the
oscillation parameters
were obtained  on
the basis of helium-overproduction, caused by the kinetic effects of 
oscillations~\cite{NPS}.

In this work,  accounting for all the  effects (a) and (b), we calculate 
$Y_p(\delta N_s, \delta m^2, sin^22\theta)$   for  $0<\delta N_s \le 1$ values,  
  and reveal  the dependence of
 $\delta N_{kin}$ on  $\delta N_s$.

We have analyzed  the self-consistent evolution of the oscillating neutrino 
and the nucleons from the neutrino decoupling epoch at $\sim 2$ 
MeV till the 
freezing of nucleons. We have followed the line of work described in 
detail in ref.~\cite{res},    
 omitting the assumption for negligible density 
of the sterile neutrinos at the onset of   $\nu_e\leftrightarrow
\nu_s$ oscillations.   

It is hardly possible to describe analytically, without some radical
approximations, the
non-equilibrium picture of
active--sterile neutrino oscillations,  producing non-equilibrium
neutrino number densities and  distorting the  neutrino spectrum.
 Satisfactory precise analytical description was found only for 
the case of relatively fast oscillations  proceding before neutrino 
freezing, with $\delta m^2>10^{-6}$ eV$^2$ and small mixing 
angles~\cite{dol2003, Villante}.
Therefore, we have provided a self-consistent numerical analysis of the 
evolution of 
the
nucleons number densities  
 $n_n$ and the ones of the  oscillating neutrinos  $\rho$ and $\bar{\rho}$  
 in the high-temperature Universe, using  
 the following coupled integro-differential equations, 
the first equation describing the kinetics of the neutrino ensembles in
terms of the density matrix of neutrino $\rho$ and anti-neutrino 
$\bar{\rho}$, the second
equation --  the kinetic evolution of the neutrons.
 
\begin{eqnarray}
&&{\partial \rho(t) \over \partial t} =
H p_\nu~ {\partial \rho(t) \over \partial p_\nu} +
\nonumber\\
&&+ i \left[ {\cal H}_o, \rho(t) \right]
+i \sqrt{2} G_F \left(\pm {\cal L} - Q/M_W^2 \right)N_\gamma
\left[ \alpha, \rho(t) \right]
+ {\rm O}\left(G_F^2 \right),
\label{kin}
\end{eqnarray}
\begin{eqnarray}
&&\left(\partial n_n / \partial t \right)
 = H p_n~ \left(\partial n_n / \partial p_n \right) +
\nonumber\\
&& + \int {\rm d}\Omega(e^-,p,\nu) |{\cal A}(e^- p\to\nu n)|^2
\left[n_{e^-} n_p (1-\rho_{LL}) - n_n \rho_{LL} (1-n_{e^-})\right]
\nonumber\\
&& - \int {\rm d}\Omega(e^+,p,\tilde{\nu}) |{\cal A}(e^+n\to
p\tilde{\nu})|^2
\left[n_{e^+} n_n (1-\bar{\rho}_{LL}) - n_p \bar{\rho}_{LL}
(1-n_{e^+})\right].
\end{eqnarray}
where $\alpha_{ij}=U^*_{ie} U_{je}$,
$p_\nu$ is the momentum of electron neutrino,
 $n$ stands for the number density of the interacting particles,
${\rm d}\Omega(i,j,k)$ is a phase-space factor, and  ${\cal A}$ is the
amplitude of the corresponding process.
The plus sign  in front of ${\cal L}$ corresponds to the neutrino
ensemble, the  minus sign - to the anti-neutrino ensemble.

Mixing just in the electron sector is assumed: 
$\nu_i=U_{il}~\nu_l$ ($l=e,s$).
The initial condition for the 
neutrino ensembles in the interaction basis
is of the form:
$$
{\cal \rho} = n_{\nu}^{eq}
\left( \begin{array}{cc}
1 & 0 \\
0 & S
\end{array} \right),
$$
where $n_{\nu}^{eq}=\exp(-E_{\nu}/T)/(1+\exp(-E_{\nu}/T))$, while $S$ 
measures the degree of population of the sterile state. 

${\cal H}_o$ is the free neutrino Hamiltonian.
The `non-local' term $Q$ arises as a $W/Z$ propagator effect,
$Q \sim E_\nu~T$.
${\cal L}$ is proportional to the fermion asymmetry of the plasma
and is essentially expressed through the neutrino asymmetries
${\cal L} \sim 2L_{\nu_e}+L_{\nu_\mu}+L_{\nu_\tau}$,
where
$L_{\mu,\tau} \sim (N_{\mu,\tau}-N_{\bar{\mu},\bar{\tau}})/ N_\gamma$
and $L_{\nu_e} \sim \int {\rm d}^3p (\rho_{LL}-\bar{\rho}_{LL})/N_\gamma$.

The equations are for the neutrino and neutron number densities {\it in
momentum space}, which  allows  to describe precisely  the kinetic 
effects: spectrum
distortion and neutrino-antineutrino asymmetry growth  due to oscillations.
For the description of the spectrum 1000 bins were used in the nonresonant 
oscillations case, and at least 5000 in the resonant one.   
These equations   provide a simultaneous account of the different
competing processes,
namely: neutrino oscillations, Universe expansion, neutrino forward 
scattering,  nucleons transformations. 

The analysis was performed  for all mixing angles
$\vartheta$ and
 mass differences
$\delta m^2 \le 10^{-7}$ eV$^2$.
The analyzed temperature interval was $[2.0, 0.3]$ MeV, because 
at  temperatures higher than $2$ MeV the
deviations from the standard BBN model without oscillations are
negligible in the discussed model of oscillations. 

As expected, the spectrum distortion is less expressed when increasing
the degree of population of the sterile neutrino state $\delta N_s$ (Fig.1.).
Correspondingly, the kinetic effect
on primordial nucleosynthesis should decrease.
The results of our numerical analysis on spectrum distortion at different 
$\delta N_s$ are illustrated in the  Figs. 1a-c, where   
 the dependence of the energy spectrum distortion of the 
electron neutrino on the initial population of the sterile state is shown. 

At each temperature  we have plotted  the spectrum 
for three different levels of initial population of the sterile neutrino, namely
$\delta N_s=0.0, 0.5, 0.8$.

\mbox{\hspace{2cm}}\epsfig{figure=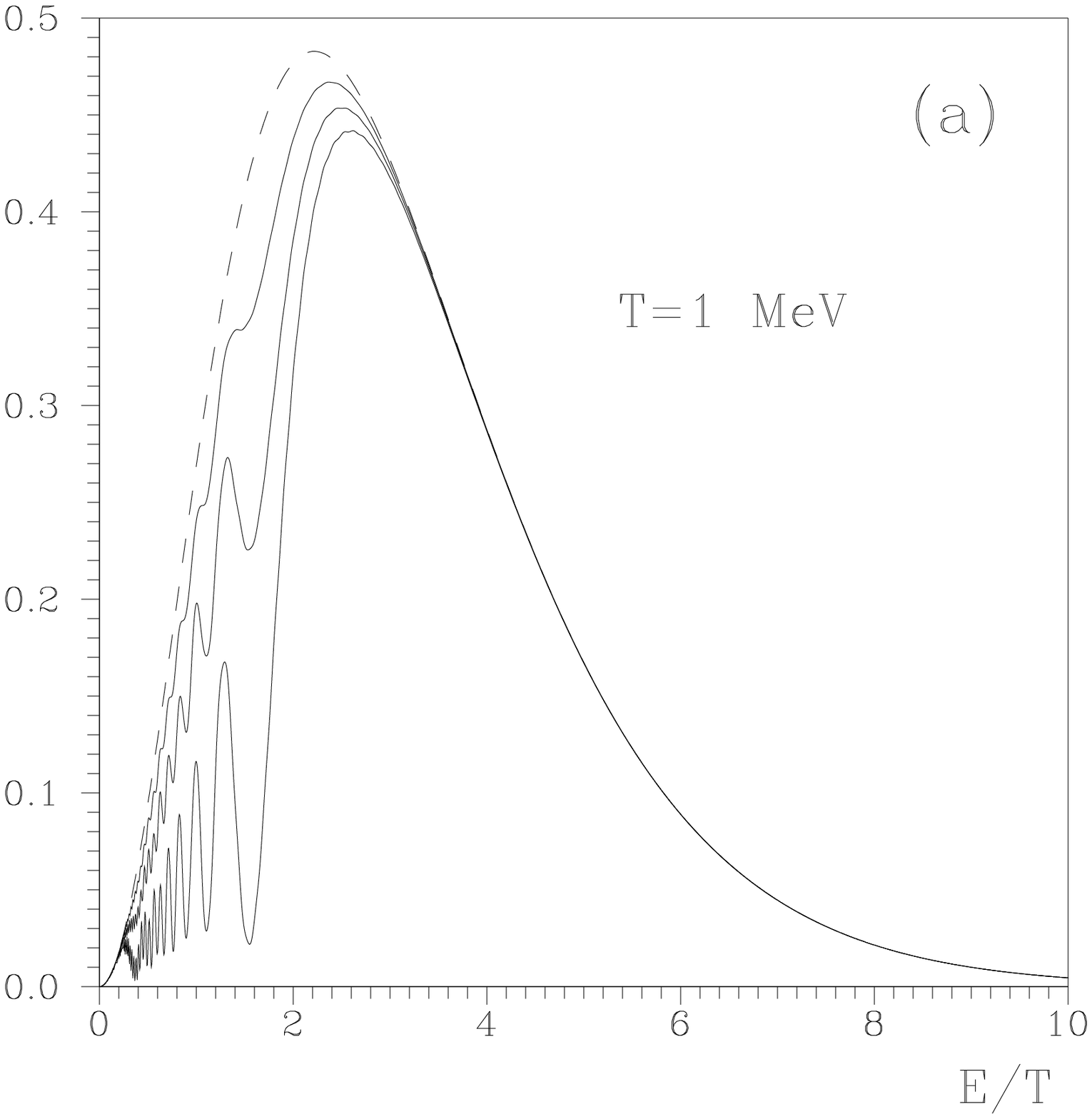,width=10cm}\\
{\bf Figure 1a:} {\small The figure illustrates the degree of 
distortion
of the electron neutrino energy spectrum $x^2\rho_{LL}(x)$, where
$x=E/T$  at a characteristic temperature  $1$ MeV,
caused by resonant oscillations with a  mass difference
$\delta m^2=10^{-7}$ eV$^2$ and $\sin^22\vartheta=0.1$
for different  initial sterile neutrino populations, correspondingly
 $\delta N_s=0$ (the lower curve),  $\delta N_s=0.5$
 and  $\delta N_s=0.8$ (the upper curve). The
dashed curve gives the
equilibrium neutrino spectrum for comparison.}
\ \\

\mbox{\hspace{2cm}}\epsfig{figure=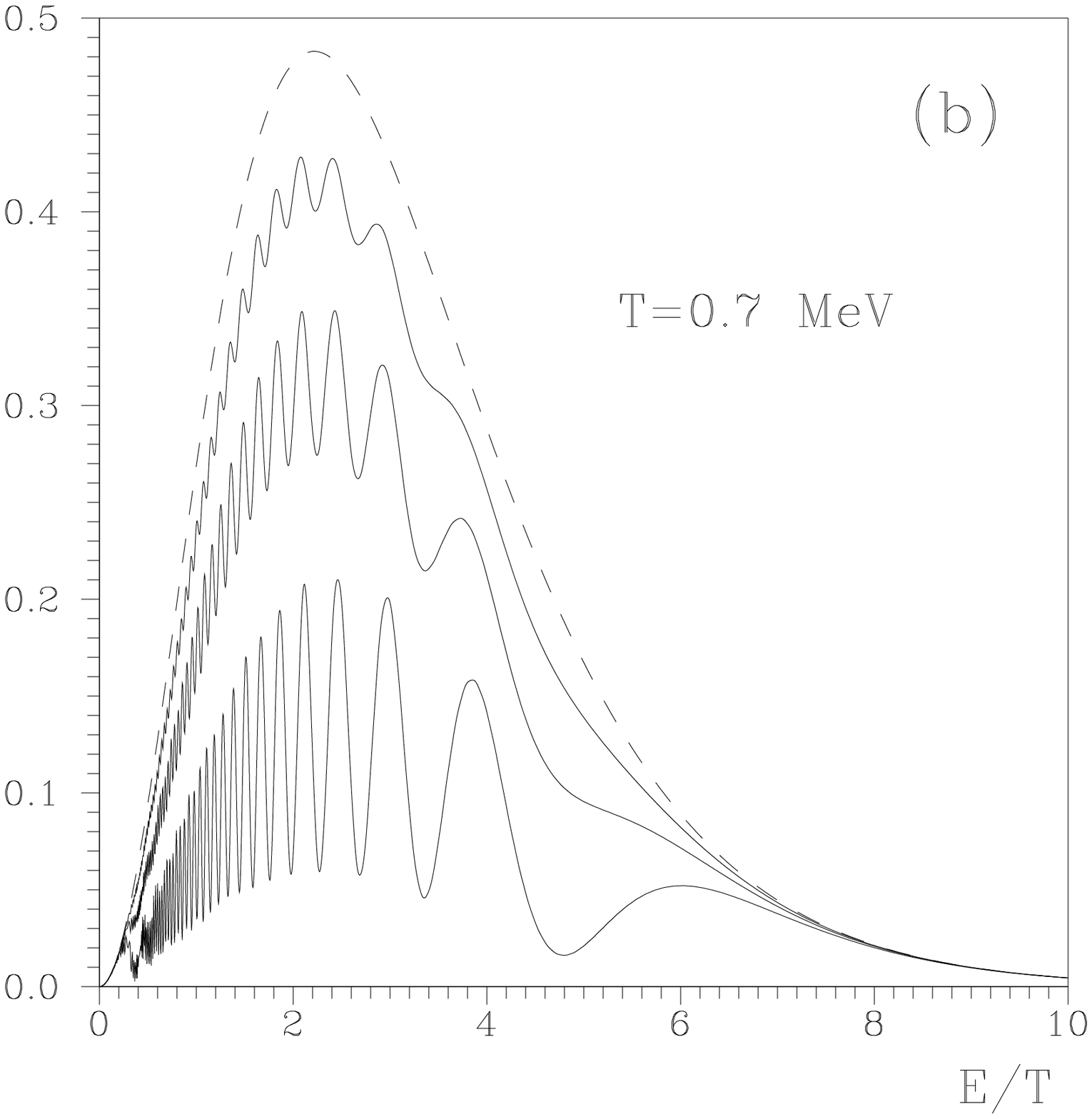,width=10cm}\\

\hbox{\vspace{-1cm}}
\mbox{\hspace{2cm}}\epsfig{figure=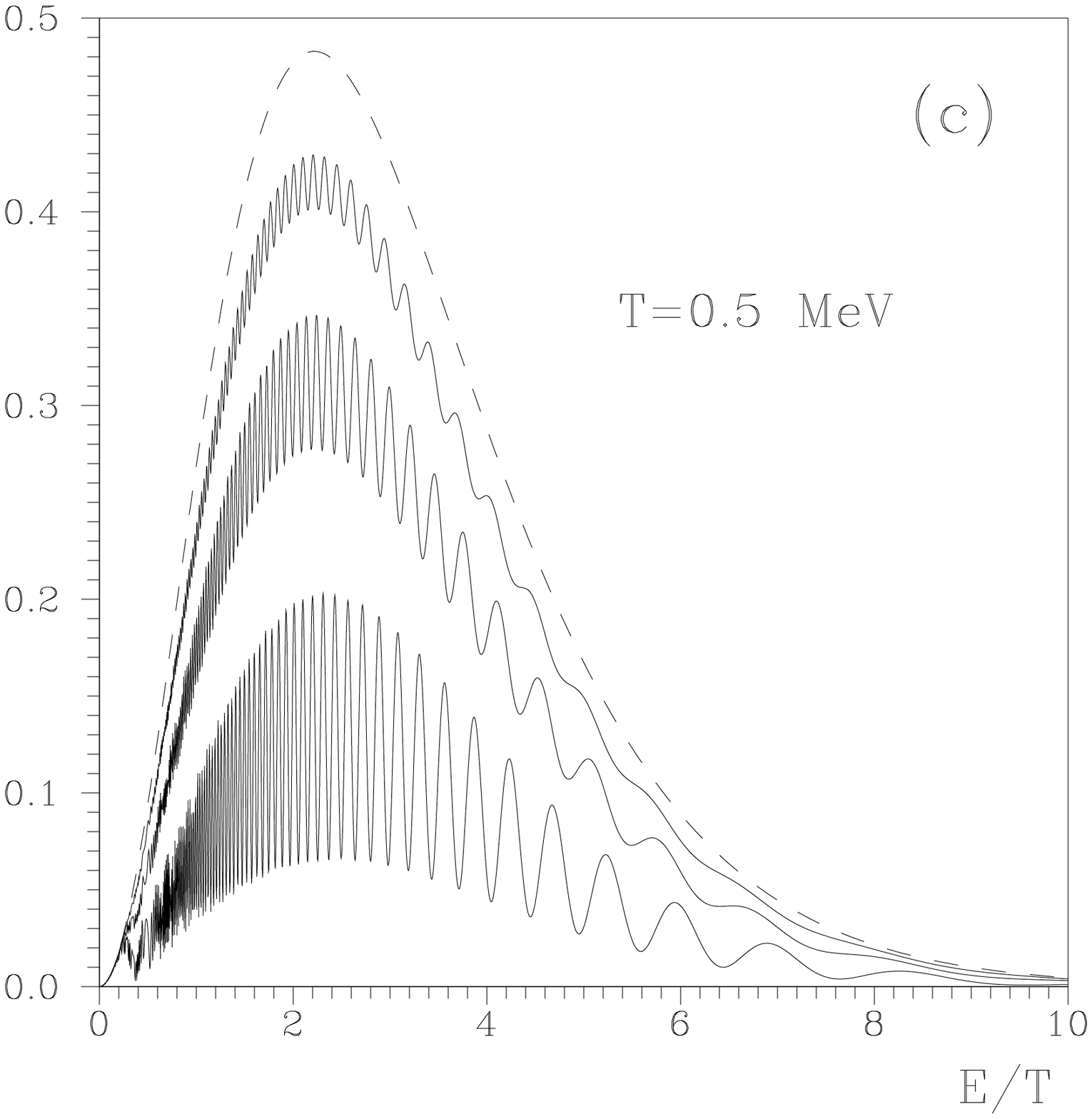,width=10cm}\\
{\bf Figures 1b,c:} {\small Distortion
of the electron neutrino energy spectrum at a temperature $0.7$ MeV
(Fig.1b) and  $0.5$ MeV (Fig.1c) for the same
parameters as for Fig.1a}.

The oscillations parameters are
$|\delta m^2|=10^{-7}$ eV$^2$ and $\sin^22\vartheta=0.1$.
For illustrating  the
evolution of the spectrum distortion we have presented it at  
characteristic
temperatures  $1$ (Fig.1a),  $0.7$ (Fig.1b) and  $0.5$ MeV (Fig.1c).
At each $\delta N_s$ the characteristic behavior of the spectrum distortion due to 
oscillations is 
observed. Namely, since
oscillation rate is energy dependent $\Gamma \sim \delta m^2/E$ the low energy
part of the spectrum is distorted first (as far as low energy neutrinos 
start to oscillate first) 
and later the
distortion  penetrates noticeably into  the more energetic part of the spectrum. 

We have found that the neutrino energy spectrum $n_{\nu}(E)$ may strongly deviate
from its equilibrium form during all the period of interest 
($2$ MeV -- $0.3$ MeV) even for considerably large  $\delta N_s$,  
and hence the spectrum distortion  may constitute the dominant effect on 
the 
overproduction of  
$^4\!$He. 

\section*{The kinetic effect}

For different  $\delta N_s$ we calculate precisely
the $n/p$-freezing, essential for the
production of helium, down to temperature $0.3$ MeV. Then we calculate $Y_p$,  
accounting  adiabatically
for the following decay of neutrons till the start of nuclear reactions,
at about $0.1$ MeV.

We have found that {\it neutrino spectrum distortion effect on BBN is
very strong  even when there is a  considerable
population of the sterile neutrino state  before the
beginning of the electron--sterile oscillations.}
It always gives positive $\delta N_{kin}$, which for a large range of
initial sterile population values, are  bigger than $1$. 
 The kinetic effects are
the  strongest for $\delta N_s=0$:
 $Y_p^{max}(\delta N_s,\delta m^2,\sin^22\vartheta)$ $=$
$Y_p(0,\delta m^2,\sin^22\vartheta)$.
They disappear for $\delta N_s=1$, when
$\nu_e$ and  $\nu_s$
states are in equilibrium, and the total effect reduces to the SBBN with
an additional neutrino.

In Fig.2 we present  the frozen  neutron number density relative to 
nucleons $X_n^f=N_n^f/N_{nuc}$    as a function of
the sterile neutrino content  at neutrino decoupling for a resonant and a nonresonant 
oscillation case.  
The oscillation parameters are  $\delta m^2=10^{-7}$ eV$^2$ and 
$\delta m^2=-10^{-7}$ 
eV$^2$ 
and  $\sin^2 2\theta=10^{-1}$.  
As far as $\delta Y_p/Y_p=\delta X_n^f/X_n^f$,  the results are  representative for   
the overproduction of 
 primordially produced helium. 

\mbox{\hspace{1cm}}\epsfig{figure=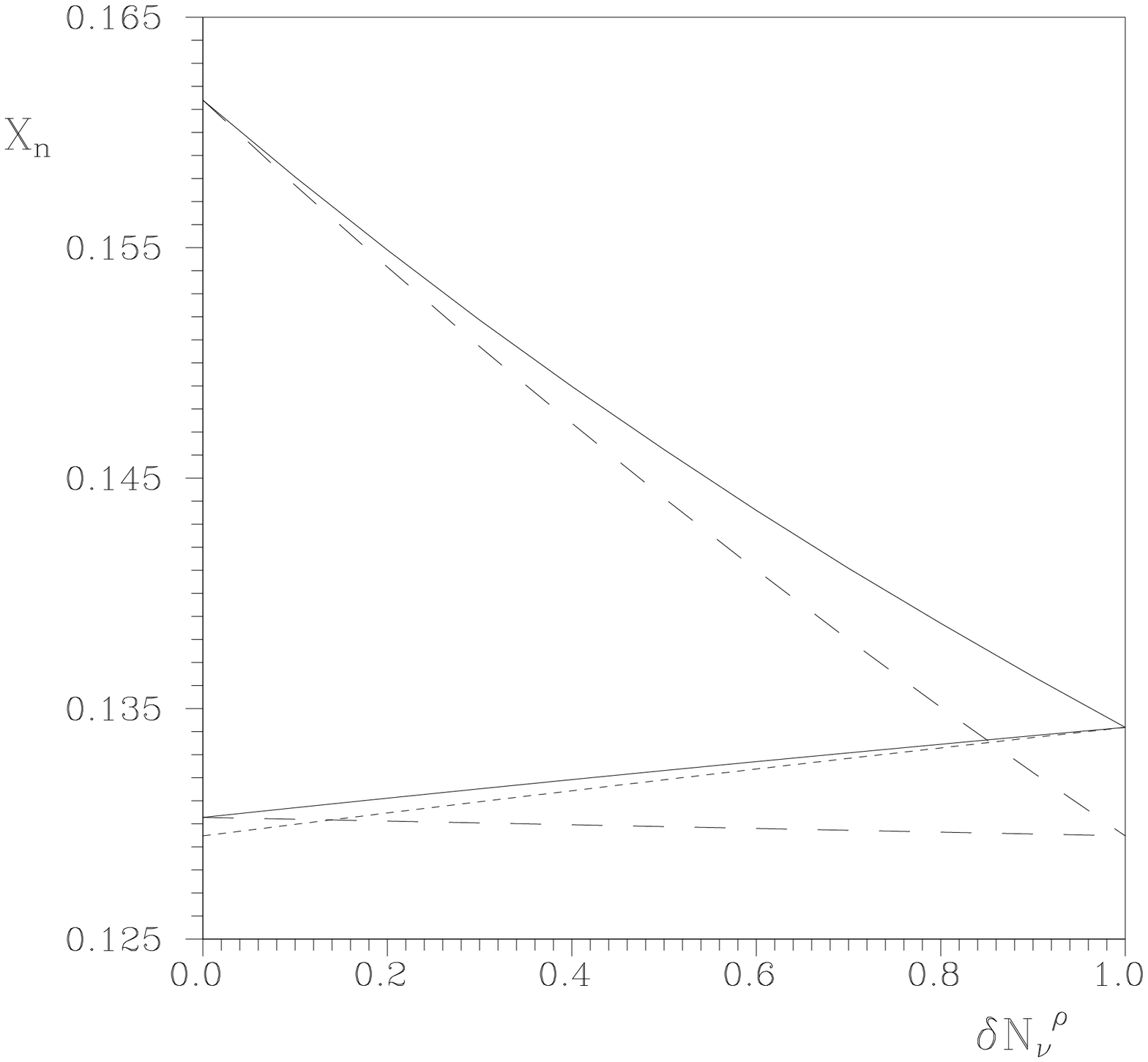,width=10cm}\\
{\bf Figure 2: }{\small The  solid curves present frozen  
neutron number density relative to
nucleons $X_n^f=N_n^f/N_{nuc}$ as a function  of
the sterile neutrino initial population. The dashed curves present only the 
kinetic effect, while the dotted curve presents the effect due to the 
energy 
density increase. The upper two curves (dashed and solid) correspond to the resonant 
case, the lower dashed and solid curves - to the nonresonant one.}
\ \\

The dotted curve  presents only the  effect (a), due to the energy 
density
increase $X_n^f=f(\delta N_s)$, the dashed curves present the pure 
kinetic
effects (b)  $X_n^f=f(\delta N_{kin})$,
 while  the solid lines give the total effect.
The upper dashed and solid curves correspond to the
resonant case, the  lower ones to the non-resonant one.

The analysis for these concrete oscillation parameters, shows that the 
overproduction
of helium is strongly 
suppressed with the increase of 
$\delta N_s$ for the resonant case, while in the non-resonant case it 
increases with 
$\delta N_s$. This is a result of the fact that,  
in the resonant 
case, the kinetic effects (b) due to the spectrum distortion are the 
dominant contribution to the overproduction of helium,   
 even for very large degree of population of the sterile state, while in 
the non-resonant case the main contribution  
comes from the increase of degrees of freedom already at very small 
$\delta N_s$. 
An  empirical approximation formula   
is: 
\begin{center}
$\delta Y_p = 0.013[\delta N_{kin}^{max}(1-\delta N_s)+\delta N_s]$,\\
\end{center}
\noindent  where $\delta N_{kin}^{max}$ is the value calculated in the 
case of oscillations with an 
initially  empty sterile state,  
i.e. $\delta N_{tot}=\delta N_{kin}^{max}(1-\delta N_s)+\delta N_s$. 
It is a good approximation for the non-resonant case and a  rather rough 
one for the resonant case: the deviation from  the exactly 
calculated helium given in Fig.~2 may be up to $\delta Y_p/Y_p\sim 0.8\%$. 
Still, it can give some idea of $\delta Y_p/Y_p$ dependence 
on $\delta N_s$.

However, for other  mixing parameters, 
 the kinetic oscillation effects in the 
non-resonant 
case can be also considerable, as shown in ref.\cite{bern}, the kinetic 
effect can be as high as $\delta N_{kin} \sim 3$ for initially empty sterile 
state. Hence, in the non-resonant case the spectrum distortion effects may be 
the dominant one even for much larger  $\delta N_s$ than in the case illustrated 
in Fig.2. 

For each  concrete  $\delta N_s$ value  a detailed numerical analysis
is necessary to reveal the interplay of  effects (a) and (b) and their influence 
on primordial production of $^4\!$He. 

In a forthcoming paper we apply the results obtained here to define 
  the isohelium  contours
corresponding to  $3\%$ overproduction of  $^4\!$He 
 for different   $\delta N_s$, and we present  the 
cosmological constraints for nonzero  $\delta N_s$. 

\section*{Conclusions}

The presence of a {\it  non-empty} sterile  
state before   $\nu_e\leftrightarrow \nu_s$  oscillations was not considered in
previous
analysis of   $\nu_e\leftrightarrow \nu_s$ oscillation effects on the 
neutrino
spectrum
distortion and  on
BBN. In this work we have studied the kinetic effects due to 
 $\nu_e\leftrightarrow \nu_s$  oscillations in the general case $0 \le \delta N_s \le 
1$.

We have provided a numerical analysis,  investigating
how the presence of the sterile neutrino state, partially populated before  
oscillations, will influence  the production of  $^4\!$He in the model
of BBN with electron--sterile oscillations effective after electron neutrino 
decoupling. 

{\it We have found that the effect of the neutrino spectrum distortion  due to 
oscillations may 
be
very strong,  even for a  considerable initial
population of the sterile neutrino state.} Correspondingly, 
the kinetic effect of 
oscillations 
remain the dominant one even for big  $\delta N_s$. 

The results of this analysis may be applied  for 
different models generating  sterile neutrino, like GUT models, mirror models, 
extra-dimensions models, etc., as far as the initial value of  population of the sterile  
state $\delta N_s$  depends on the concrete model of its production.
These results may be of interest also for 
mixing  schemes in which a 
portion of $\nu_s$ have been brought into equilibrium before 
neutrino decoupling, due to  $\nu_{\mu}\leftrightarrow \nu_s$ or
$\nu_{\tau}\leftrightarrow \nu_s$ oscillations.
In case the $\nu_s$ presence is
due to the much earlier (at atmospheric mass difference scale, or LSND)
oscillations of $\nu_{\mu,\tau}\leftrightarrow\nu_s$,
 $\delta N_s$  may be  directly connected with the available constraints
on the sterile
neutrino fraction, deduced
from the neutrino oscillations experimental data analysis.
So, we hope that the results   may be indicative and helpful for choosing  among 
the different  possibilities for the sterile fraction in the  
subdominant active-sterile oscillations used in the oscillation 
analysis of  
neutrino anomalies.  

A 
more general  study  of kinetic  oscillations effects on BBN 
 for non-empty initially  
 sterile state  in the framework of  4-neutrino mixing
schemes seems appropriate, 
although much more complicated. 

\ \\

I thank A. Dolgov for useful comments on the draft version, 
M. Chizhov and A. Strumia for stimulating discussions during
the preparation of this work. I am 
grateful to my parents for their help with the kids, during the  work  on
this  paper.

I appreciate the  Corresponding Associate position at CERN TH and the 
Regular  Associateship  at the Abdus Salam ICTP, Trieste.
This work  was supported in part by Belgian Federal Government (office for 
scientific affairs grant and IAP 5/27).

\end{document}